\documentclass[preprint2]{aastex631}
\shorttitle{\textit{INTEGRAL}/SPI Study of MAXI J\(1535-571\)}
\shortauthors{Rodi et al.}
\graphicspath{{./}{figures/}}

\begin{document}

\title{MAXI J\(1535-571\) 2017 outburst Seen by \textit{INTEGRAL}/SPI and Investigating the Origin of Its Hard Tail}

\correspondingauthor{James Rodi}
\email{james.rodi@inaf.it}

\author[0000-0003-2126-5908]{James Rodi}
\affiliation{INAF - Istituto di Astrofisica e Planetologia Spaziali; via Fosso del Cavaliere 100; 00133 Roma, Italy}

\author[0000-0001-9932-3288]{E. Jourdain}
\affiliation{Universit\'e de Toulouse; UPS-OMP; IRAP;  Toulouse, France}
\affiliation{CNRS; IRAP; 9 Av. Colonel Roche, BP 44346, F-31028 Toulouse cedex 4, France}

\author[0000-0002-1757-9560]{J. P. Roques}
\affiliation{Universit\'e de Toulouse; UPS-OMP; IRAP;  Toulouse, France}
\affiliation{CNRS; IRAP; 9 Av. Colonel Roche, BP 44346, F-31028 Toulouse cedex 4, France}



\begin{abstract}
On 2 September 2017 MAXI J\(1535-571\) went into outburst and peaked at \( \sim \) 5 Crab in the \(2-20\) keV energy range.  Early in the flare \textit{INTEGRAL} performed Target of Opportunity pointings and monitored the source as it transitioned from the hard state to the soft state.  Using quasi-simultaneous observations from \textit{MAXI}/GSC and \textit{INTEGRAL}/SPI, we studied the temporal and spectral evolution of MAXI J\(1535-571\) in the \(2  - 500\) keV range.  Early spectra show a Comptonized spectrum and a high-energy component dominant above \( \sim 150\) keV.  \texttt{CompTT} fits to the SPI data found electron temperatures (\(kT_e \)) evolves from \( \sim 31\) keV to 18 keV with a tied optical depth (\( \tau \sim 0.85\)) or \( \tau\) evolving from \( \sim 1.2 - 0.65\) with a tied \( kT_e \) (\( \sim 24\) keV).  To investigate the nature of the high-energy component, we performed a spectral decomposition of the \(100-400\) keV energy band.  The \texttt{CompTT} flux varies significantly during the hard state while the high-energy component flux is consistent with a constant flux.  This result suggests that the two components originate from different locations, which favors a jet origin interpretation for the high-energy component over a hybrid corona interpretation.  Lastly, two short rebrightenings during the hard-to-soft transition are compared to similar events reported in MAXI J\(1820+070\).

\end{abstract}


\keywords{X-rays: general --- X-rays: binaries --- Black holes: individual (MAXI J\(1535-571\))} 


\section{Introduction} \label{sec:intro}

The transient black hole candidate (BHC) MAXI J\(1535-571\) went into outburst on 2 September 2017 \citep{negoro2017} and remained active for over a year with short reflaring events after \(\sim 200\) days.  Due to its high intensity and long duration, the source was a frequent target from radio to X-rays.  

To date, most results have focused on the X-ray and radio behavior of MAXI J\(1535-571\) with works from \textit{AstroSat} \citep{sridhar2019}, \textit{Insight}-HXMT \citep{kong2020},  MAXI/GSC \citep{nakahira2018},  NICER \citep{miller2018,cuneo2020}, \textit{NuSTAR} \citep{xu2018}, and \textit{Swift} \citep{tao2018} reporting on the X-ray spectral characteristics and spectral state transitions during various parts of the outburst and reflaring.  

Additional X-ray analyses have investigated the timing behavior and found type-A, B, and C quasi-periodic oscillations (QPOs) at various points \citep{huang2018,steven2018,stiele2018,bhargava2019,mereminskiy2018,chaterjee2021}.

Results from radio have mainly explored the transient behavior of the jet and/or the relationship between the jet and accretion disk \citep{parikh2019,russell2019,russell2020a,chauhan2021}.  \cite{chauhan2019} used radio observations to determine the distance to the source (\(\sim 4\) kpc).

Here we propose to focus on the soft gamma-ray (\( \gtrsim 100 \) keV) behavior.  The origin of this emission in BHs and BHCs in the hard state is debated, but it can be phenomenologically described as a power-law or a cutoff power-law model and is observed above the thermal Comptonization component.  Proposed mechanisms include bulk-Comptonization \citep{laurent1999}, Comptonization of non-thermal electrons \citep{wardzinski2001}, possibly in a hybrid thermal/non-thermal plasma \citep{coppi1999}, and (synchrotron) jet emission \citep{laurent2011}.

This high-energy component, sometimes called a ``hard tail,'' has been observed in both persistent (Cyg X-1 \citep{jourdain1994,cadollebel2006} and 1E \(1740.7-2942\) \citep{bouchet2009}) and transient (e.g. V404 Cyg \citep{jourdain2017}, MAXI J\(1820+070\) \citep{roques2019}, GRS \(1716-249\) \citep{bassi2020}) BH(C)s.

MAXI J\(1535-571\) provides an unusual opportunity to study the hard tail, because the source intensity allowed studying the behavior on the timescale of a few hours, and also to observe how the evolution compares to other wavelengths, in particular as the source  transitioned from the hard to soft state.  Such a comparison is usually difficult as long exposure times are often required to detect hard tails or the state transition is not observed with instruments sensitive to soft gamma-rays.

\begin{table*}
\begin{center}
\caption{\textit{INTEGRAL}/SPI observations of MAXI J\(1535-571\)}
\scalebox{0.925}{
\hspace{-20mm}
\begin{tabular}{ccccc}
\tableline\tableline \\
\textit{INTEGRAL}  & Time                                           & Time                 & SPI Exp.  & MAXI/GSC Exp.\\
Revolution         & (UTC)                                          & (MJD)                &    (ks)            &    (ks)\\  
\tableline
1860                         & 2017-09-08 12:46:01 \(-\) 2017-09-10 16:52:24  & 58004.5319 \(-\) 58006.7030 &  116.6  &  6.9  \\ 
1862                         & 2017-09-13 20:42:21 \(-\) 2017-09-16 01:42:37  & 58009.8627 \(-\) 58012.0712 &  76.7   &  6.3 \\ 
1863                         & 2017-09-16 12:31:36 \(-\) 2017-09-18 17:33:00  & 58012.5219 \(-\) 58014.7312 &  115.4  &  5.2  \\
1864                         & 2017-09-19 07:12:51 \(-\) 2017-09-21 03:37:02  & 58015.3005 \(-\) 58017.1507 &  130.0  &  3.5  \\
1865                         & 2017-09-21 20:09:54 \(-\) 2017-09-23 23:19:22  & 58017.8402 \(-\) 58019.9717 &  115.4  &  2.2 \\
\tableline
\end{tabular}
}
\label{tab:obs}
\end{center}
\end{table*}

\section{Instrument and Observations}
The \textit{INTErnational Gamma-Ray Astrophysics Laboratory} (\textit{INTEGRAL}) was launched on 17 October 2002 from Baikonur, Kazachstan with an eccentric \( \sim 3\)-day orbital period \citep{jensen2003}.  The SPectrometer on-board \textit{INTEGRAL} (SPI) covers the \(20\) keV \(-\) 8 MeV energy range and has an energy resolution of \( 2-8 \) keV \citep{roques2003}.  \textit{INTEGRAL} began Target of Opportunity observations of MAXI J\(1535-571\) during \textit{INTEGRAL} revolutions \(1860-1865\) (2017-09-08 12:46:01 \(-\) 2017-09-23 23:19:22 UTC).
Observations during revolution 1861 and a part of revolution 1862 suffered from solar activity and thus could not be used.  The remaining data during this period were analyzed using the SPI Data Analysis Interface (SPIDAI)\footnote{Publicly available interface developed at IRAP to analyze SPI data. Available at http://sigma-2.cesr.fr/integral/spidai. See description in \citet{burke2014}} to construct light curves and spectra. The background determination (detector plane uniformity pattern) is based on empty field observations, temporally close from the source observation, with an amplitude adjusted on a \(3-11\) hr timescale. The spectra are binned in 35 energy channels in the \(25-500\) keV.

The Monitor of All-sky X-ray Image (MAXI) is an experiment onboard the International Space Station (ISS) \citep{matsuoka2009}.  The Gas Slit Camera (GSC) instrument covers the \(2-30\) keV energy range and can observe \( \sim 85\%\) of the sky during an ISS orbit.  Due to its large field-of-view, the GSC detected MAXI J\(1535-571\) on 2 September 2017 \citep{negoro2017} and was able to follow the evolution of the source throughout its outburst.  However, here we focused on GSC observations during \textit{INTEGRAL} revolutions \(1860-1865\).  The spectra were generated using the online MAXI/GSC spectral extraction tool\footnote{http://maxi.riken.jp/mxondem/} using the default parameters and were binned in \(192-302\) channels from \(2-20\) keV depending on the number of counts during the observation.  The light curve data are from the publicly available light curves\footnote{http://maxi.riken.jp/top/slist.html}. 

All of the spectra for both instruments have at least 10 counts per bin and thus can be analyzed using \(\chi^2\)-statistics \citep{cash1979}.  Time ranges and exposure times for the SPI and MAXI/GSC observations are listed in Table~\ref{tab:obs}.

\begin{figure}[h!]
  \centering
  \includegraphics[scale=0.65, angle=0,trim = 15mm 10mm 20mm 100mm, clip]{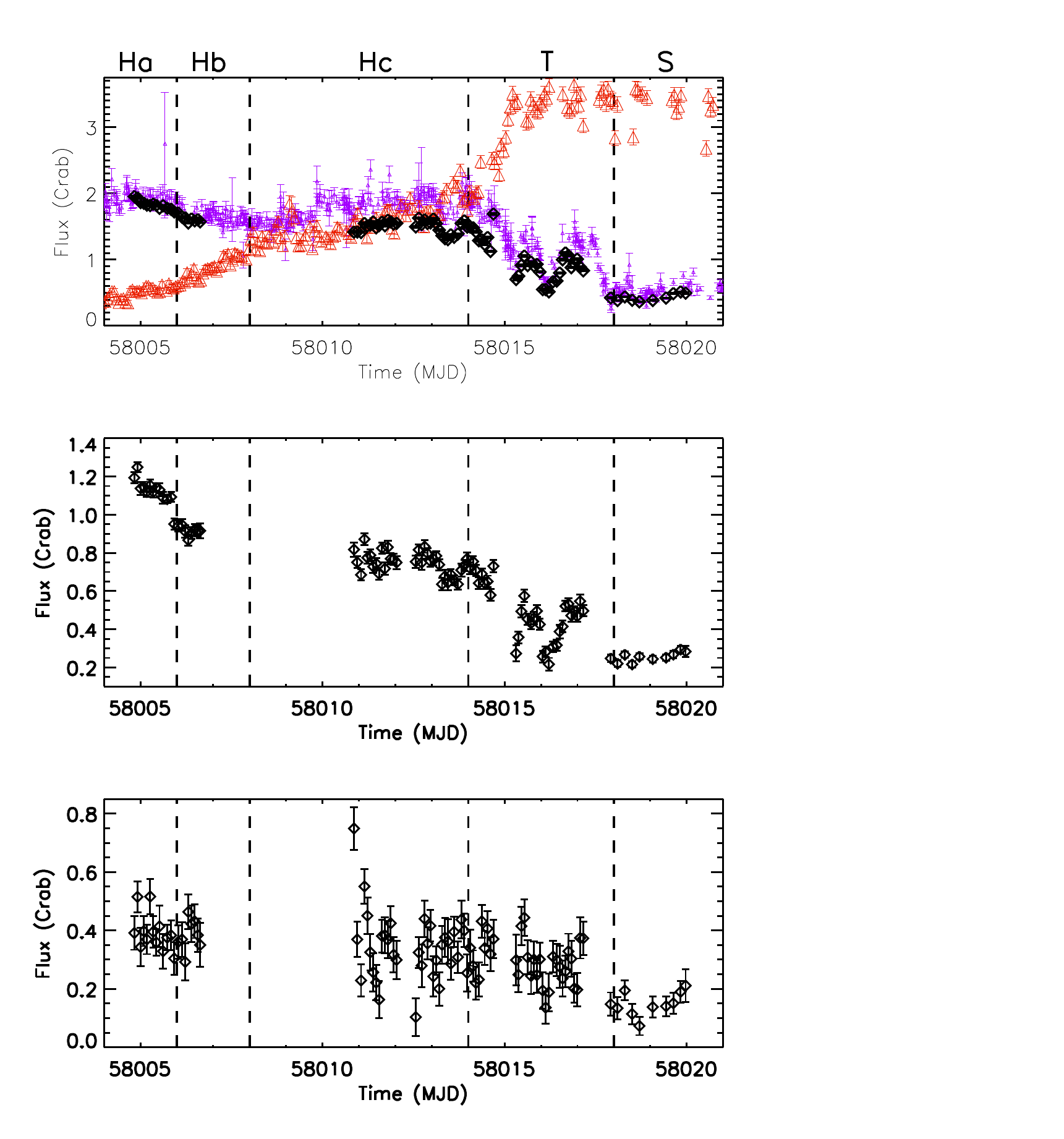}
  \caption{\textit{Top}: MAXI/GSC (\(2-4\) keV; red triangles) and \textit{Swift}/BAT (\(15-50\) keV; purple squares) orbital light curves, together with the \(25-50\) keV SPI (black diamonds) light curve on a 2-scw timescale until \(\approx \) MJD 58017, after which the light curve is on a 5-scw timescale.  The SPI \(50-100\) keV \textit{(Middle)} and \(100-400\) keV \textit{(Bottom)} light curves are show using the same time binning.  The vertical dashed lines in each panel denote state changes based on \citep{nakahira2018} (see text).  Quasi-simultaneous MAXI/GSC and SPI observations have been used to study the spectral behavior of the source.} \label{fig:lc}
\end{figure}

\begin{figure}[h!]
  \centering
  \includegraphics[scale=0.65, angle=0,trim = 15mm 10mm 20mm 100mm, clip]{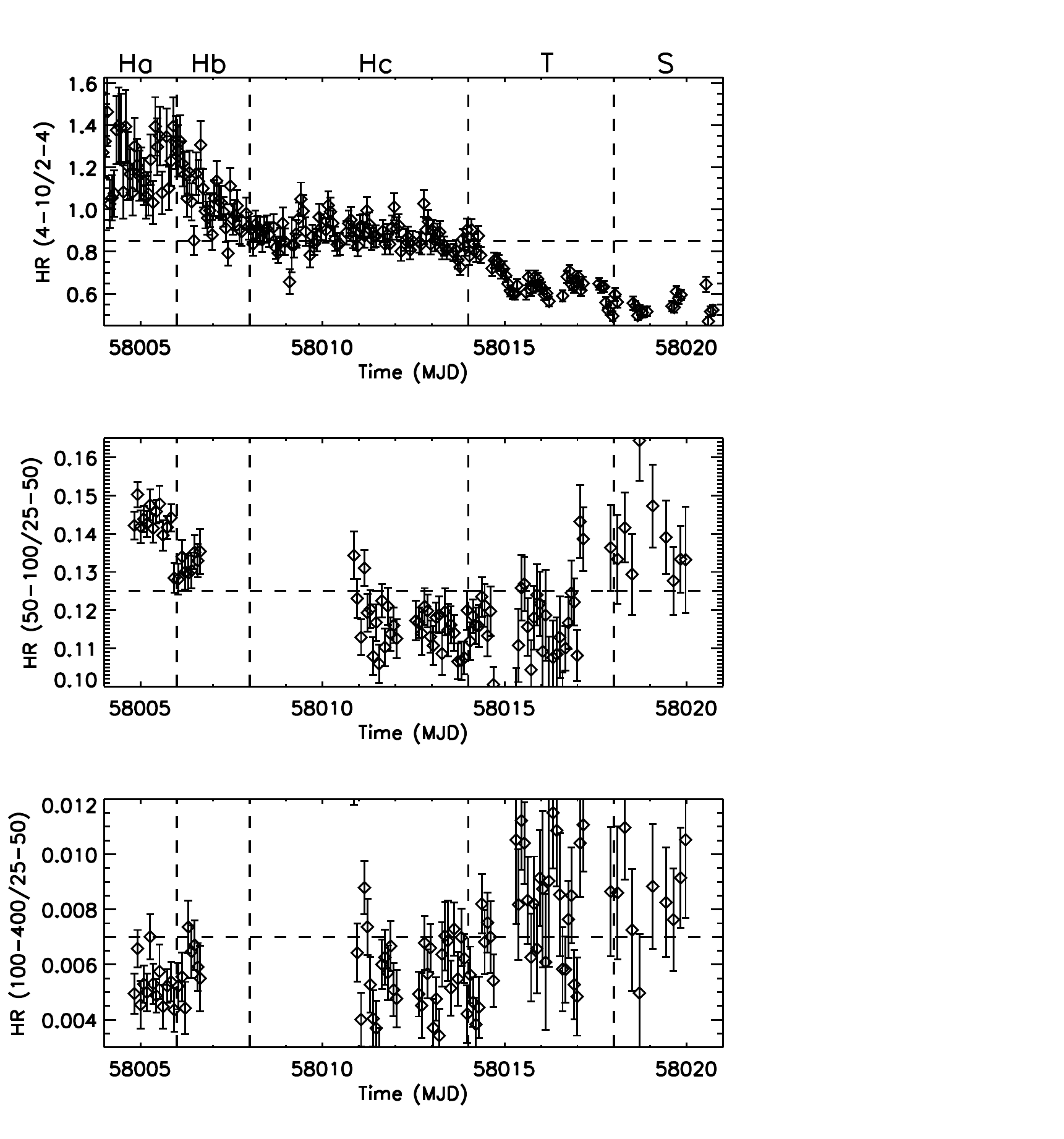}
  \caption{\textit{Top}: Time history of GSC (\(4-10\) keV)/(\(2-4\) keV) hardness ratios.  The vertical dashed lines in each panel denote state changes based on \citep{nakahira2018}. \textit{Middle}:  Time history of SPI (\(50-100\) keV)/(\(25-50\) keV) hardness ratios.  \textit{Bottom}: Time history of SPI (\(100-400\) keV)/(\(25-50\) keV) hardness ratios.  The dashed horizontal lines in each panel are to guide the eye.} \label{fig:hr}
\end{figure}

\section{Analysis and Results}

\subsection{Temporal Behavior}
MAXI J\(1535-571\) was first detected at 23:55 UT 2 September 2017 \citep{negoro2017}.  The \textit{INTEGRAL} ToO began roughly 5 days later.  During the first observations, SPI observed the source at approximately 2 Crab in the \(25-50\) keV band.  The top panel of Figure~\ref{fig:lc} shows the \(25-50\) keV light curve in black diamonds on a 2-science window (scw) timescale (\( \sim 1800\) s per science window) until \( \sim\) MJD 58018 then a 5-scw timescale, with the orbital timescale light curves of \textit{MAXI}/GSC (\(2-4\) keV) and \textit{Swift}/BAT (\(15-50\) keV) overplotted in red triangles and purple squares, respectively.  The \(1 \sigma\) errors are shown in the figure.  The dashed vertical lines mark changes in the spectral states based on \cite{nakahira2018}.  Periods `Ha',`Hb', and `Hc' are times when the source was in a hard state, but differing behaviors in the hardness ratios.  The `S' period is when the source was in a soft state, and the `T' period denotes to a transitional period between the hard and soft states.

We recall that the different spectral states are related to different accretion states in the system and the relative contributions of thermal emission from an accretion disk and Comptonized emission from a corona, which up-scatters photons from the accretion disk (\cite{remillard2006} and references within).  The geometry of the corona is an open question though.  In the hard state (HS), the X-ray spectrum is dominated by the Comptonization component.  In contrast, the thermal component dominates during the soft state (SS).  Between the two states are intermediate states as the system transitions to the hard and soft states and vice versa.

The different states of the source are reflected in the behavior of its high-energy emission:
From the beginning of the observations until \(\sim\) MJD 58015 (i.e. during the hard states), the soft (below \(\sim 10\) keV) and hard (above \(\sim 15\) keV X-ray bands evolve independently with the soft X-rays showing a gradual increase and the hard X-rays with a long, shallow dip. During the transition, (i.e. after MJD 58015), the two bands show an anti-correlation with a sharp increase in the soft X-rays followed by a plateau with a corresponding sharp decrease in the hard X-rays with two rebrightenings during MJD \(58015-58018\) before settling to a roughly, constant flux.  The first rebrightening lasts approximately 0.5 days, and the second lasts roughly 2 days.  During both these features, the soft X-ray flux is nearly constant. At the end of the \textit{INTEGRAL} observations, the soft state is established, with a low hard X-ray flux and a strong emission in soft X-rays.

\begin{figure}[h!]
  \centering
  \includegraphics[scale=0.70, angle=0,trim = 0mm 115mm 50mm 45mm, clip]{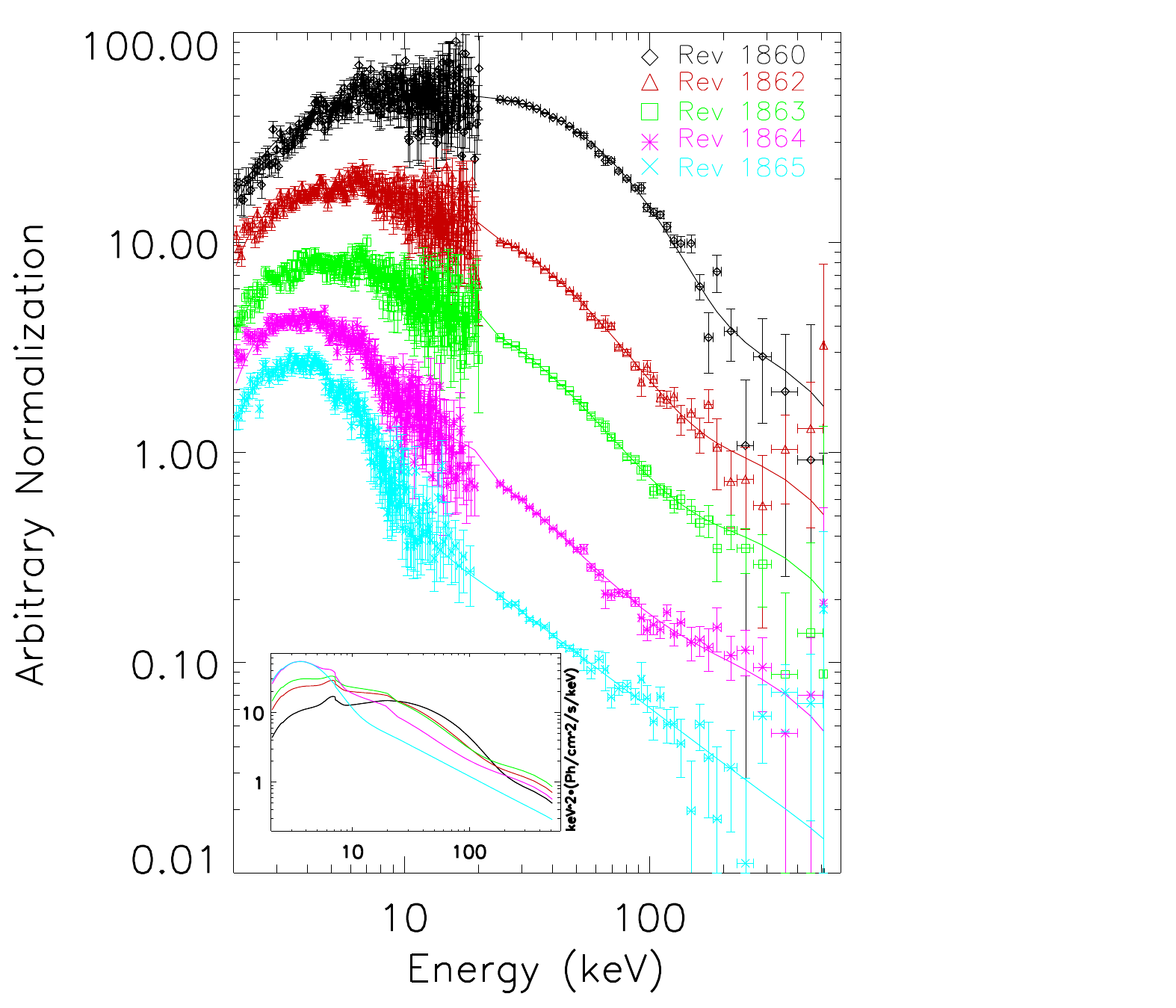}
  \caption{The average spectra for revolutions 1860, 1862\(-\)1865 in the \(2-500\) keV energy range arbitrarily scaled.  The model spectra are shown in the inset without scaling in units of keV\(^2\)*(photons/cm\(^2\)/s/keV).  For revolutions 1860, 1862, and 1863, the model is \texttt{Tbabs*Refl*(CompTT)+Cutoff}.  For revolution 1864, the model is \texttt{Tbabs*Refl*(diskbb+CompTT)+Cutoff} and for revolution 1865, the model is \texttt{Tbabs*(diskbb+po)}. }
\label{fig:spec_avgs}
\end{figure}

The subsequent panels in Figure~\ref{fig:hr} show the evolution of the hardness ratios (HR) in the \(4-10\) keV\(/\)\(2-4\) keV, \(50-100\) keV\(/\)\(25-50\) keV, and \(100-400\) keV/\(25-50\) keV (respectively). Dashed lines have been added to guide the eye at 0.85, 0.12, and 0.007, respectively.  The vertical dashed lines are again the state transition from \cite{nakahira2018}.  The first portion of both the \(4-10\) keV \(/\) \(2-4\) keV and \(50-100\) keV \(/\) \(25-50\) keV HR time histories show a gradual softening that lasts until \( \sim\) MJD 58008, based on the GSC data.  In contrast, the \(100-400\) keV \(/\) \(25-50\) keV HR time history shows the HR are fairly constant during this period.  

The MJD \(58008-58014\) period, preceding the transition to the soft state, is characterised by all the considered hardnesses roughly constant. Around MJD 58015, the softening of the X-ray emission is accompanied by a hardening of the E \(> 100\) keV emission, while the  \(25-100\) keV hardness remains unchanged and will increase 3 days later. After that, HR stay constant again, but with a configuration opposite to that observed before the transition.

\begin{figure}[t!]
  \centering
  \includegraphics[scale=0.68, angle=0,trim = 25mm 0mm 10mm 170mm, clip]{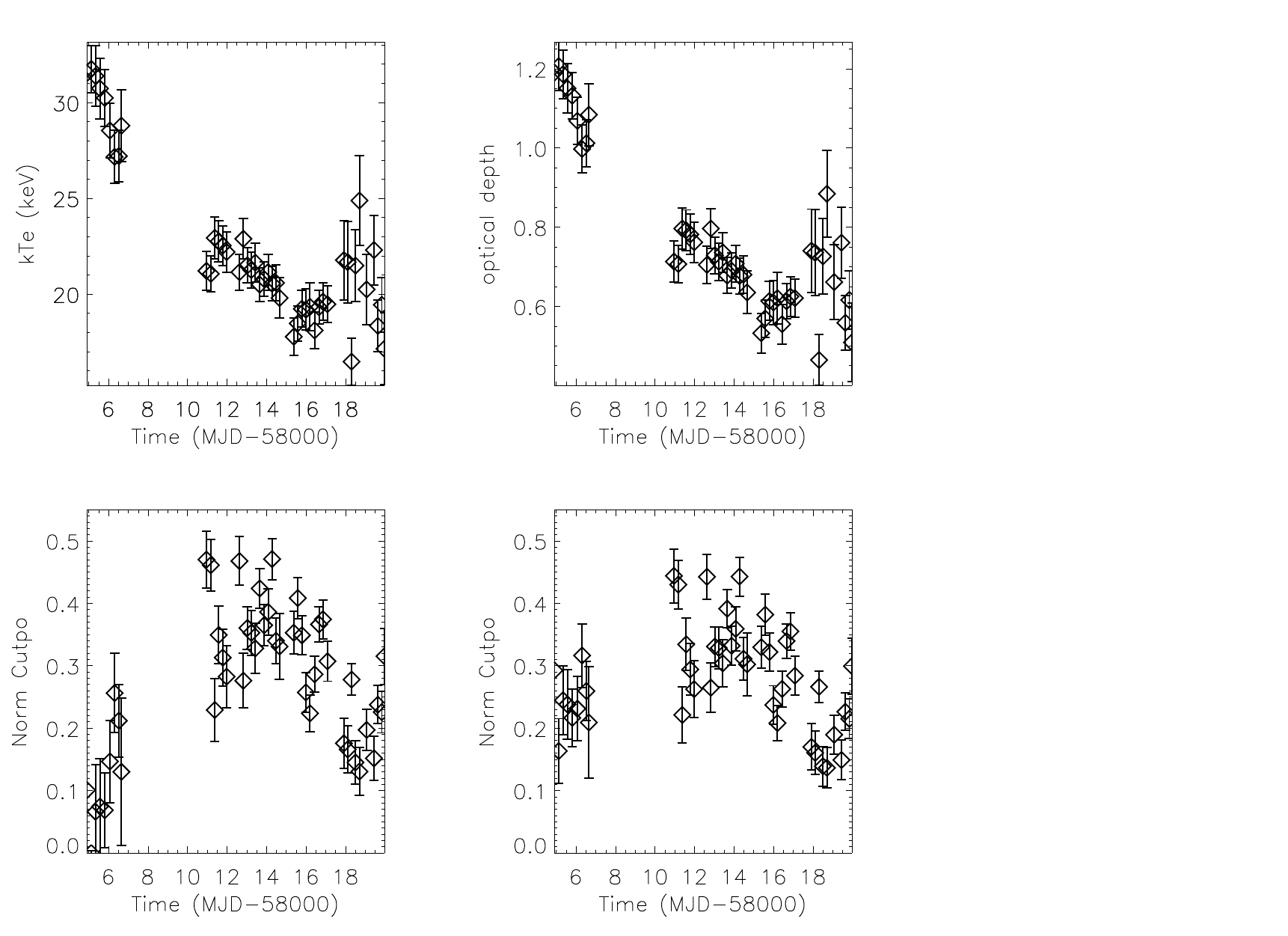}
  \caption{\textit{Top}: Evolution of optical depth (right) during revolutions 1860, \(1862-1865\) with \(kT_e = 23.9 \pm 0.09\) keV and evolution of electron temperature (left) during revolutions 1860, \(1862-1865\) with \( \tau = 0.85 \pm 0.05 \).  \textit{Bottom}: Evolution of cutoff power-law normalization during revolutions 1860, \(1862-1865\) for a variable electron temperature and fixed optical depth (left) and for a fixed electron temperature and variable optical depth (right).} 
\label{fig:specvar}
\end{figure}

\subsection{Spectral Behavior}
\subsubsection{Revolution Variability}
To start, we carried out fitting of the joint GSC/SPI \(2-500\) keV spectra on a revolution timescale using \texttt{Xspec}.  All the best-fit parameter errors  are given at the \(1-\sigma\) level.  Based on MAXI/GSC results from \citep{nakahira2018}, we used \(N_{H} = 2.6 \times 10^{22} \textrm{ cm}^{-2} \) and assumed abundances and photoionization cross-sections from \cite{wilms2000}.

Absorbed power law (\texttt{const*tbabs*po}) models for revolutions 1860, 1862, and 1863 (all corresponding to the source in the hard state) found photon indexes of \(2.718 \pm 0.005\), \(2.806 \pm 0.008\), and \(2.896 \pm 0.007\), respectively.  The models were poor fits to the data with \(\chi^2/ \nu\) values of \(4415.70/327 = 13.5\),  \(2738.76/334 = 8.2\), and \(2210.53/326=6.8\).  In each fit, the model overpredicts the flux \(\sim < 4\) keV and \(\sim > 50\) keV.  The high-energy residuals suggest a cutoff, indicative of Comptonization.

Fitting the spectra to an absorbed cutoff power-law model (\texttt{const*tbabs*cutoff}) in Xspec, with fixed $N_{H}$) better described the data with \(\chi^2/ \nu\) values of \(443.71/324 = 1.37\) (null hypothesis probability\footnote{In \texttt{Xspec} the null hypothesis is that the observed data are drawn from the model.}  (NHP) \(= 1.1 \times 10^{-6}\)), \(511.45/333 = 1.53\) (NHP \(= 1.1 \times 10^{-9}\)), and \(542.01/325 = 1.67\) (NHP \(= 4.3 \times 10^{-13}\)) for revolutions 1860, 1862, and 1863, respectively.  The best-fit parameters for each revolution are \( \Gamma = 1.78 \pm 0.02\) and \(E_c = 49.6 \pm 0.09\) keV, \( \Gamma = 2.20 \pm 0.01\) and \(E_c = 56.1 \pm 1\) keV, and \( \Gamma = 2.38 \pm 0.01\) and \(E_c = 71 \pm 2\) keV for revolutions 1860, 1862, and 1863, respectively.  The difference in \( \chi^2\) values between the power-law and cutoff power-law models gives an estimate of the significance of the cutoff.  A comparison between the models finds \( \Delta \chi^2 \) values of 3971.99, 2227.31, and 1668.52 for revolutions 1860, 1862, an 1863, respectively. However, the models underpredict the flux \(\sim > 100\) keV, which suggests the presence of an additional component above the \texttt{cutoff} model.

Next, the hard state spectra were fit with a more physically motivated model using a \texttt{CompTT} model that was convolved with a reflection model \citep{magdziarz1995} (assuming abundances from \cite{anders1989}) and a broad Fe line, which were reported in \cite{sridhar2019} and \cite{kong2020}.  The reflection factor was fixed to 1, indicating a disk geometry, and the energy and width of the line were fixed to 6.5 keV and 0.8 keV, respectively.  The seed photon temperature (\(kT_0\)) was fixed to 0.4 keV  \citep{kong2020}.  The \(\chi^2/ \nu\) values for each fit were \(450.61/323 = 1.40\) (NHP \(= 3.3 \times 10^{-6}\)), \(454.47/332 = 1.37\) (NHP \(=8.8 \times 10^{-6}\)), and \(445.07/324 = 1.37\) (NHP \(= 8.9 \times 10^{-6}\)).  As in the cutoff power-law model, high-energy excesses above \( \sim 100\) keV are seen in the residuals of each spectrum.


Following \cite{roques2019}, we tested the presence of a high energy tail  by adding a cutoff power-law with a photon index fixed to 1.6 and a cutoff energy fixed to 300 keV, leaving only the normalization free.  Including this component further improved the quality of fit for each of the spectra with \( \chi^2 / \nu = 421.41/322 = 1.31\) (NHP \(= 1.6 \times 10^{-4}\)) for revolution 1860, \( \chi^2 / \nu = 392.01/331 = 1.18\) (NHP \(= 1.2 \times 10^{-2}\)) for revolution 1862, and \( \chi^2 / \nu = 367.24/323 = 1.14\) (NHP \(= 4.5 \times 10^{-2}\)) for revolution 1863.  We note that revolution 1860 shows the most significant intra-revolution spectral evolution in the HR plots, which may explain a poor description obtained for the average spectrum.  

To investigate the probability of improvement by chance, we performed an F-test.  For revolution 1860, the probability was \(3.5 \times 10^{-6}\), for revolution 1862 the probability was \(2.7 \times 10^{-12}\), and for revolution 1863 the probability was \(3.5 \times 10^{-15}\).  Thus including the cutoff component is statistically justified. The best-fit parameters corresponding to this model are given in Table~\ref{tab:paras}, together with the cross-calibration factor for the MAXI/GSC instrument relatively to SPI (constant factor fixed to 1).

From MJD 58014 to 58018, \cite{nakahira2018} found MAXI J\(1535-571\) in clearly a transition between HS and SS, thus disk emission rises during revolution 1864. So a \texttt{diskbb} model was included in an absorbed power-law model to account for the emission below \( \sim 10\) keV.  This results in best-fit parameters values of \(kT_{in} = 1.61 \pm 0.03\) keV and \( \Gamma = 3.02 \pm 0.01\) with  \(\chi^2 / \nu = 306.95/277 = 1.11\) (NHP \(= 1.0 \times 10^{-1}\)).  Though the fit is acceptable, the residuals in the SPI energy range (above \(\sim 50\) keV) are similar to those observed in the power-law fits in the hard state revolutions.  A fit to the SPI data alone using an absorbed power-law model finds a similar photon index (\(3.03 \pm 0.01\)), but with \(\chi^2 / \nu = 56.03/33 = 1.70\) (NHP \(= 7.4 \times 10^{-3}\)), thus revealing that the SPI data are not well-described by a power-law.  We thus fit the revolution 1864 to the same model as revolutions 1860, 1862, and 1863 but with a \texttt{diskbb} component with the diskbb temperature tied to the seed photon temperature.  The best-fit parameters are reported in Table~\ref{tab:paras}.  However, the source evolution during this revolution appears complex with rebrightenings seen in the SPI light curves (Fig.~\ref{fig:lc}) and variability in the MAXI/GSC hardness ratios (Fig.~\ref{fig:hr}), as well as changes in the timing behavior \citep{huang2018,stiele2018} during this period.  This limits any conclusion from the analysis of the average spectrum on an \textit{INTEGRAL} revolution timescale (\(\sim 2\) days).  In later analysis, we investigate the spectral variability on a shorter timescale. 

Finally, during revolution 1865, where the source is in a soft state \citep{nakahira2018}, the spectrum is well-described by an absorbed \texttt{diskbb+po} model with \( \chi^2 / \nu =193.93/221 = 0.88\) (NHP \(= 9.1 \times 10^{-1}\))  and flat residuals.  The best-fit parameters are reported in Table~\ref{tab:paras} as for the previous revolutions.

\begin{table*}
\begin{center}
\caption{Revolution Timescale Spectral Parameters}
\hspace{-30mm}
\scalebox{0.95}{
\hspace{-6mm}
\begin{tabular}{cccccccccc}
\tableline \\
\multicolumn{1}{c}{}     & \multicolumn{6}{c}{\texttt{Const*Tbabs*Refl*(Diskbb+CompTT)+Cutoff\footnote{For the cutoff power-law \(\Gamma =\) 1.6 and \(E_c =\) 300 keV.} }}  & \multicolumn{3}{c}{\texttt{Const*Tbabs*(Diskbb+Powerlaw)}}\\
\tableline
Rev.   & Cross-Normalization & \(kT_{in}/kT_0\)        & \(kT_e\)        &   \(\tau\)          &  & \(\chi^2 / \nu\) & \(kT_{in}\) &  \(\Gamma\) & \(\chi^2 / \nu\) \\
       &                     & (keV)              & (keV)           &                     &                  &            & (keV)      &                  \\
\tableline                   \\
1860   &  \(1.05 \pm 0.02\)  &  0.4 (fixed)       & \(21.4 \pm 0.9\)&  \(1.31 \pm 0.06\)  &  &  1.31 & \(-\)            &\(-\)              & \(-\)\\ 
1862   &  \(1.12 \pm 0.03\)  &  0.4 (fixed)       & \(22 \pm 2\)    &  \(0.85 \pm 0.09\)  &  &  1.18 & \(-\)            &\(-\)              & \(-\)\\ 
1863   &  \(1.18 \pm 0.03\)  &  0.4 (fixed)       & \(28 \pm 2\)    &  \(0.55 \pm 0.07\)  &  &  1.14 & \(-\)            &\(-\)              & \(-\)\\
1864   &  \(1.22 \pm 0.08\)  &  \(0.84 \pm 0.03 \)& \(62 \pm 43\)   &  \(0.1 \pm 0.2\)    &  &  1.22 & \(-\)            &\(-\)              & \(-\)\\ 
1865   &  \(0.98 \pm 0.07\)  & \(-\)              &  \(-\)          & \(-\)               &  & \(-\) & \(1.15 \pm 0.03\)& \(2.88 \pm 0.02\) & 0.88 \\ 
\tableline
\end{tabular}
}


\label{tab:paras}
\end{center}
\end{table*}

The average spectrum for each revolution is shown in Figure~\ref{fig:spec_avgs}.  They have an arbitrary normalization for clarity, while the best fits models are displayed in the inset to better illustrate the evolution of the spectral shape during the observations.  The spectral evolution shows the decreasing significance of the Comptonization component from revolution 1860 to 1863 with the presence of a less variable high-energy component above \( \sim 100 \) keV.  The revolution 1864 consists of a disk blackbody component with a weak Comptonization component and a similar high-energy component. Only two components are required to describe the source emission at the end of SPI observations, without it being clear which one of the Comptonized or the power law  vanished.

\subsubsection{Sub-daily Variability}
To investigate more in detail the spectral variability of the Comptonized emission, we built the SPI spectra on a 5-scw  timescale (\(\sim 5\)-hr).  Then we jointly fit the SPI 5-scw spectra from revolutions 1860, 1862\(-\)1865 using the \texttt{reflect*(CompTT)+cutoff} model.  The evolution of the best-fit parameters with time are shown in Figure~\ref{fig:specvar}.  The reflection fraction was fixed to 1, the index of the cutoff component to 1.6 and the cutoff energy to 300 keV, as above. To deal with the degeneracy between the two main parameters of the Comptonization modeling, the electron temperatures (\(kT_e\)) were tied between all the spectra while the optical depths (\( \tau\)) were left free. This joint fit converges to \(kT_e = 23.9 \pm 0.9\) keV and indicates that the optical depths decrease nearly monotonically from roughly 1.2 to 0.65 with a mean error of \( \sim 0.06\) until MJD 58015 (revolution 1864) when the optical depth becomes nearly constant.   

We performed a similar analysis with \(kT_e\) free and \( \tau\) was tied.  The best-fit value for the optical depth is \(0.85 \pm 0.05\).  Like with the optical depths, we find a downward trend in electron temperature from \( \sim 32\) keV to 20 keV and a mean error value of \( \sim 1.3\) keV, with the same flattening as is seen in the changes in optical depth during the observations.  Figure~\ref{fig:specvar} shows the evolution of the free parameters (i.e  \(kT_e\) or \( \tau\), and cutoff power-law normalization) with the optical depth variability in the right panel and the electron temperature variability in the left panel.  Both scenarios produce acceptable results with \(\chi^2 / \nu\) values of \(1477.64/1394 = 1.06\) (NHP =1) and \(1519.46/1394=1.09\) (NHP =1), respectively.  Interestingly, the Compton parameters  during the rebrightenings between MJD \(58015-58017.5\) are constant.  The cutoff normalization values between both models show similar global behaviors in both scenarios with a mean error of \( \sim 0.04\).  However, the normalizations during revolution 1860 are significantly lower when the optical depths are tied with values of \( \sim 0.1\).  During the same period, the normalization values are \( \sim 0.25\) when the electron temperature is tied.

In sum, the temporal evolution of the MAXI J\(1535-571\) spectra can be thought of as the corona having a constant electron temperature (or optical depth) with a decreasing optical depth (or electron temperature) while the system transitions from the hard state to the soft state observed in revolution 1865. However, we also remind the reader that for this last revolution the flux is low and a power law is able to adequately describe those data.  Thus fitting the data with a more complex model results in parameters that are relatively poorly constrained relative to the other spectra.

\section{Discussion}
\subsection{Comparison with Other Observations}

\begin{figure}[h!]
  \centering
  \includegraphics[scale=0.7, angle=0,trim = 25mm 105mm 60mm 110mm, clip]{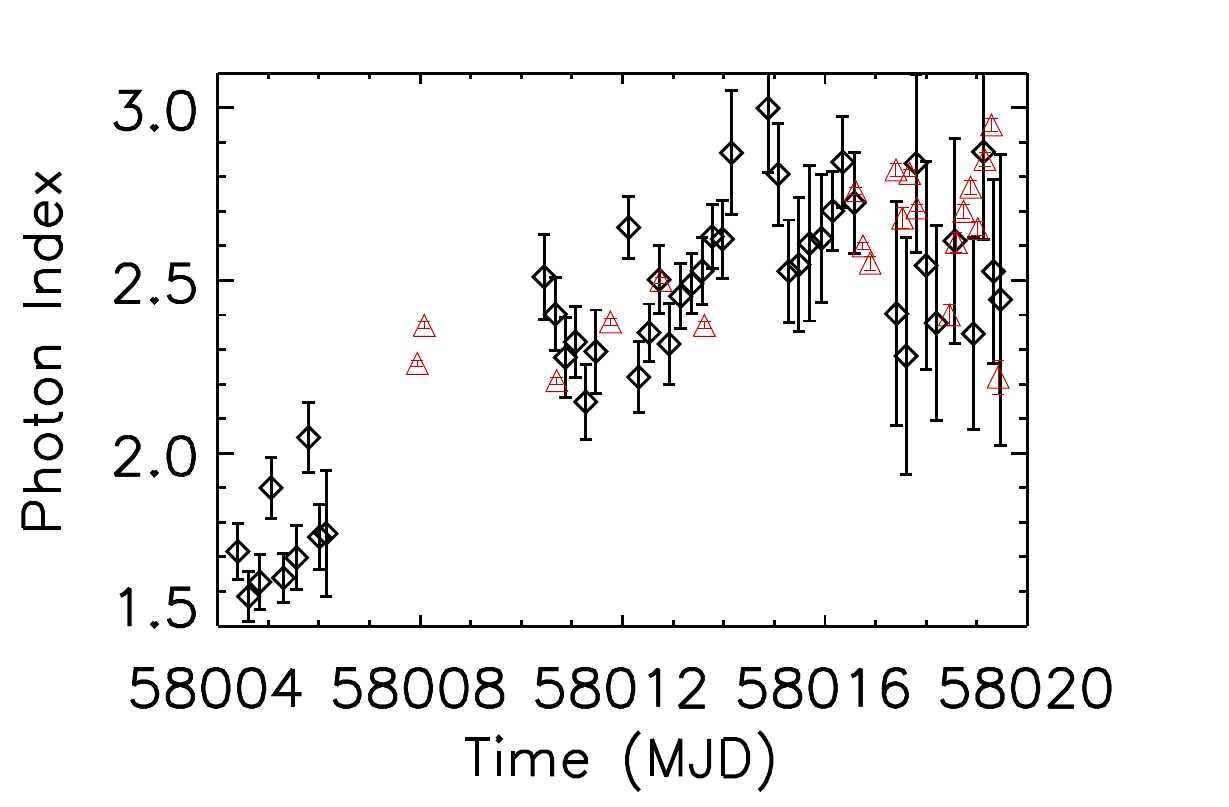}
  \caption{Evolution of photon index using a \texttt{cutoff} model to \(25-100\) keV spectra during SPI observations.  The power law indexes from \cite{kong2020} are overplotted as red triangles.} 
\label{fig:spec_po_context}
\end{figure}

\textit{Insight}/HXMT observations in the \(2-150\) keV energy range spanned MJD 58002.974 to 58019.920, overlapping with our SPI data.  \cite{kong2020} fitted their data to a \texttt{diskbb+cutoff} model and found photon indexes increasing from \( \Gamma \sim 1.5 \) at the beginning of the observations to \(\Gamma \sim 2.9\) at the end of the observations.  To compare, we fitted the SPI data in the \(25-100\) keV band on a 5-scw timescale.  As shown above (Fig~\ref{fig:spec_avgs}), SPI spectra extend to higher energies, however they show the presence of an additional component that is not seen in the \textit{Insight}/HXMT spectra.  Therefore, we did not extend up to 150 keV in the comparison as some spectra have non-negligible emission from the high-energy component at those energies.  The SPI results (Figure~\ref{fig:spec_po_context}) show a similar photon-index evolution.  In general the indexes during MJD 58018 to 58020 are lower than the HXMT values (red triangles in Figure~\ref{fig:spec_po_context}), though the errors on the SPI values are relatively large (and the cutoff energies are unconstrained), thus in agreement with \cite{kong2020}.  

\subsection{Interpretation of Spectral Evolution}

At hard X-rays the lower-energy component in the hard state is generally understood to be from the corona while the origin of the high-energy component, often referred to a a `hard tail,' is less clear.  It can be phenomenologically modeled as a power law or as cutoff power-law so it is thought to be non-thermal emission, though occasionally the component is modeled as an additional thermal Comptonization component \citep{bouchet2009}.  Early interpretations explained the high-energy emission as arising from a hybrid thermal/non-thermal Comptonizing plasma (i.e. \cite{coppi1999}).  Later interpretations explained the component as non-thermal emission related to the jet seen at radio wavelengths.   Polarization detection by \textit{INTEGRAL} in the hard tail of Cyg X-1 \citep{laurent2011,jourdain2012} strongly supports synchrotron emission from the jet.

Several models, including hybrid Comptonization, have been developed (for instance \texttt{eqpair} by \cite{coppi1999}), but they all require a large number of free parameters, resulting in degenerate parameters. These  models are not excluded by the spectral shape of MAXI J\(1535-571\), but they are not conclusive.  However, a correlation between Compton and hard tail components can be used to discriminate between the jet and hybrid Comptonization models.  If their fluxes are correlated, they likely have a common origin.  Conversely, a lack of correlation between the fluxes from the two components would suggest the photons originate from different locations.

In an attempt to differentiate between the two scenarios, we investigated the spectral decomposition of the \(100-400\) keV light curve (Figure~\ref{fig:flx_comps}).  The plot shows the \texttt{CompTT} flux (red triangles) varies significantly while the cutoff power-law flux (green squares) is nearly constant from \( \sim\) MJD \(58004-58016\).  There is a brief dip lasting \( \sim 0.5\) days (MJD \(58015.984-58016.430\)) before returning the pre-dip level until MJD 58016.852.  By MJD 58017.914 the total flux (black diamonds) drastically decreased when the SPI spectrum is just a power law. 

To quantify the variability in the cutoff power-law component prior to the dip, we fit the fluxes between MJD 58005 and 58018.7 to a constant and found a values of \(3.13 \pm 0.06 \times 10^{-9}\) erg/cm\(^2\)/s with a \( \chi^2 / \nu = 55.96/28\) (NHP \(= 1.3 \times 10^{-3}\)).  The lack of significant variability in the cutoff power-law flux prior to MJD 58016 in contrast to the \texttt{CompTT} flux evolution (Figure~\ref{fig:flx_comps}) disfavors a same origin for both components.

In the jet scenarios, the soft gamma-ray emission is thought to be produced at the base of the jet \citep{zdziarski2012}, just like some of the IR flux \citep{russell2014}, while the radio emission comes from far from the BH.  Therefore, looking at infrared observations during the hard state could provide a more relevant comparison with the hard-tail flux than radio observations.

\cite{baglio2018} found a stable IR light curves from MJD \(58002-58012\) and MJD \(58017-58029\) with a significant flux decrease between the two periods.  The spectra prior to the gap are consistent with an optically thin synchrotron emission from a jet.

On another hand, \cite{russell2019} report that the radio jet began to quench (\( \sim\) MJD 58013.5) and a subsequent likely radio flare occurred between MJD \(58013.5 - 58017.4\), complicating interpretations of the hard tail as originating from the jet.

The \(100-400\) keV cutoff power-law flux, which finds that the jet flux is not present by MJD 58017.914, is consistent with IR results from \citep{baglio2018}.  There were no observations between MJD \(58012.103\) and 58017.027 to assess when the jet emission ended or if there was a corresponding IR dip at the beginning of MJD 58016, as is seen in the \(100-400\) keV flux. 
However, the source behavior in the hard X-rays is rather chaotic between $\sim$ MJD 58015 and MJD 58018, with two rebrightenings and a peculiar spectral evolution pointed out in the hardness evolution. It may be related to the radio properties.

\begin{figure}[h!]
  \centering
  \includegraphics[scale=0.7, angle=0,trim = 8mm 10mm 50mm 214mm, clip]{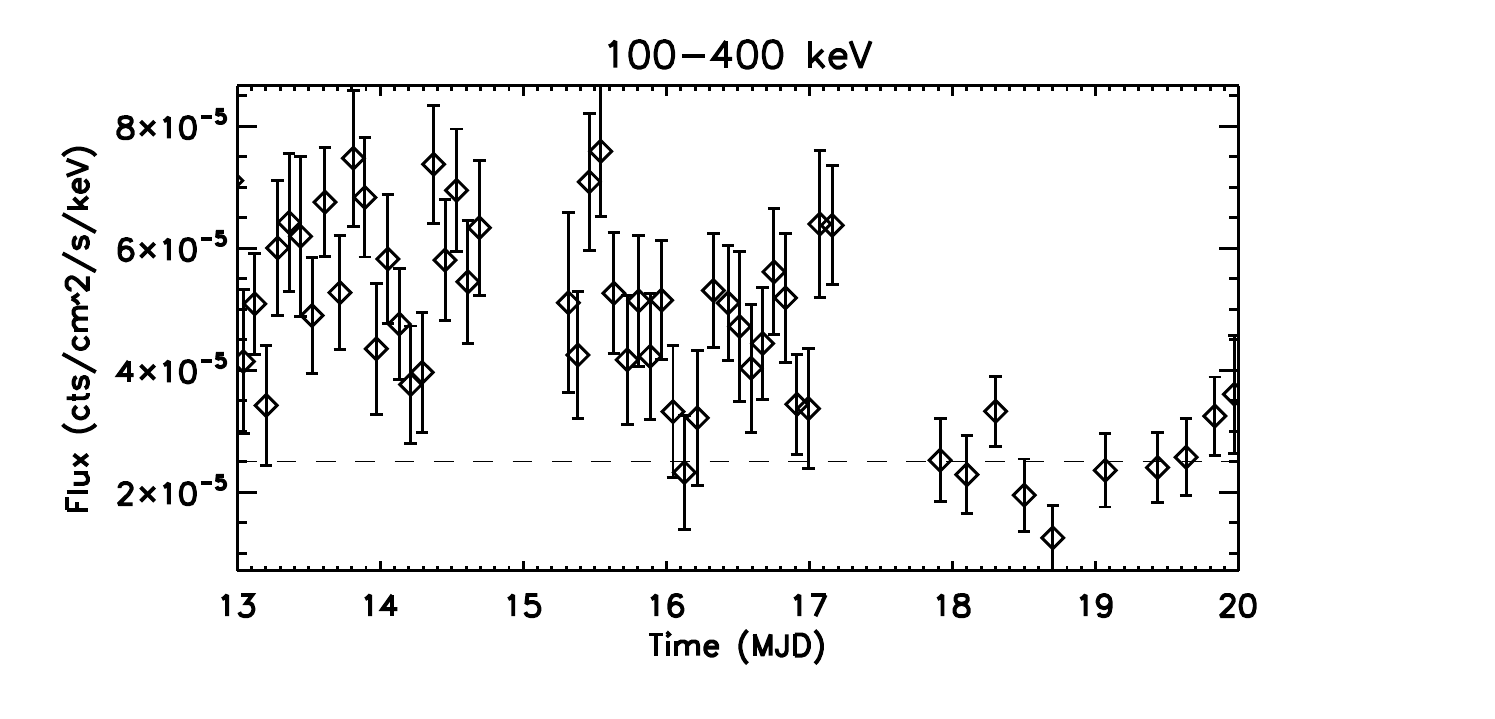}
  \caption{SPI \(100-400\) keV light curve with revolution \(1863-1864\) on a 2-scw timescale and revolution 1865 on a 5-scw timescale.  The dashed line marks the approximate level of the revolution 1865 flux.} 
\label{fig:rev1864-1865lc}
\end{figure}

Figure~\ref{fig:rev1864-1865lc} shows the \(100-400\) keV light curve for revolutions \(1863-1865\) to investigate the transition to the soft state.  Revolutions \(1863-1864\) data are on a 2-scw timescale, and the revolution 1865 data are on a 5-scw timescale. 

We see that by MJD 58016 and then again by MJD 58017, the \(100-400\) keV flux decreased to the same level as MJD \(58018-58020\), when the soft state is established.  It may suggest the disappearance / reappearance of jet activity.

Interestingly, we note that \cite{russell2020b} reported an IR flare (\( \sim 4\) days) from the BH transient 4U \(1543-47\) suggesting the presence of a compact jet, though in that case transitioned from the soft state to the SIMS before returning to the soft state.  In MAXI J\(1535-571\), the dip seen at the beginning of MJD 58016 is then perhaps better understood as a failed quenching of the jet as the nadir of the dip is at similar flux level as revolution 1865, before briefly reforming (\( \sim 0.23\) days) and re-quenching.  

\section{Comparison with MAXI J\(1820+070\)}

The BH transient MAXI J\(1820+070\) went into outburst in 2018 with \textit{INTEGRAL} observations from 16 March to 8 May (MJD \(58193-58246\)).  Due to its exceptional brightness, it triggered numerous observations and related interpretation results. Analysis of the SPI data by \cite{roques2019} in the \(25-1000\) keV energy range found the spectra underwent a similar evolution to MAXI J\(1535-571\) with the Comptonization parameters electron temperature/optical depth evolving from a high to low before flattening.  However, the MAXI J\(1820+070\) spectrum was stable for roughly 40 days compared to only a few days in MAXI J\(1535-571\).  

\begin{figure}[h!]
  \centering
  \includegraphics[scale=0.7, angle=0,trim = 28mm 10mm 60mm 155mm, clip]{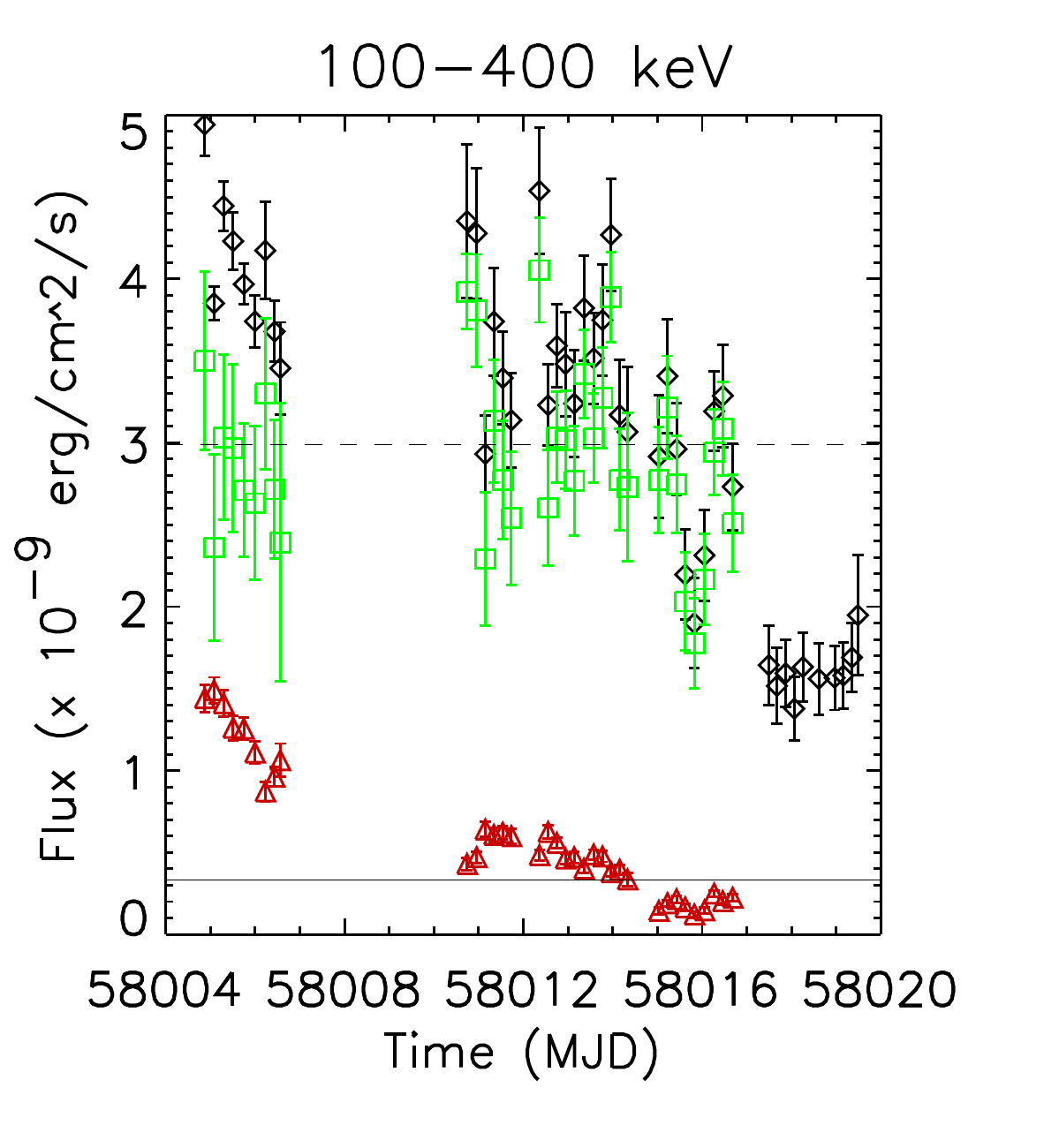}
  \caption{SPI \(100-400\) keV 5-scw timescale light curve.  The black diamonds show the total flux.  The green squares show the cutoff power-law flux.  The red triangles show the \texttt{CompTT} flux.  The best-fit model of a constant to the cutoff power-law flux is shown as a dashed line.  The best-fit model of a constant to the \texttt{CompTT} flux is shown as a solid line.} 
\label{fig:flx_comps}
\end{figure}

Another similarity is the constancy of the high-energy flux as the hard X-ray flux decreases while the source was in the hard state.  \cite{roques2019} did not perform a spectral decomposition of the high-energy flux, but they did report the cutoff power-law normalization evolution.  Their results found that the normalization stayed at a similar level from the time near the peak of the burst until the end of the \textit{INTEGRAL} observations.  In contrast, the \texttt{CompTT} normalization decreases over the same period.  Similar behavior was observed in MAXI J\(1535-571\) (Fig.~\ref{fig:flx_comps}).

MAXI J\(1820+070\) can provide some understanding of the rebrightenings in MAXI J\(1535-571\).  \citet{kara2019} interpreted early \textit{NuSTAR} observations of MAXI J\(1820+070\) as arising from a contracting corona while the inner radius of the accretion disk remained constant.  Thus the behavior in MAXI J\(1535-571\) during the transition to the soft state can be thought of as potentially an expanding and contracting corona.  

In another work, \citet{buisson2021} investigated MAXI J\(1820+070\) rebrightenings as due to disruptions of the inner accretion disk that sends material into the corona.  However, they do not propose the mechanism that causes the disruption in the disk.  Alternatively, they point out that the MAXI J\(1820+070\) rebrightening took place \( \sim 1\) day after flaring at radio wavelengths and that the rebrightening is related to the system returning to equilibrium after the ejection.  Interestingly, MAXI J\(1535-571\) had flaring at radio wavelengths before both rebrightenings.

\section{Conclusion}

We analyzed the transient BHC MAXI J\(1535-571\) in the \(2-500\) keV energy range over MJD \(58004-58020\).  During this period, the source peaked at an intensity of a few Crab, which allowed for studying the temporal and spectral variability on a timescale of \( \sim 5\) hours up to several hundred keV.  While in the hard state, MAXI J\(1535-571\) showed \(kT_e / \tau\) decreasing with time before flattening in a transition state with a hard-tail component dominant about \(\sim 150\) keV until the source moves to the soft state.  

The time evolution of the soft X-ray, hard X-ray, and soft gamma-ray hardnesses (Fig.~\ref{fig:lc}) show interesting correlations and lack of correlations during the burst with the soft and hard X-rays evolving similarly until the transition state while the soft gamma-ray hardness remains mostly unchanged.  When MAXI J\(1535-571\) softens in the soft X-rays, the hard X-ray hardness remains unchanged and the soft gamma-ray hardness increases.  Not until the source is in the soft state does the hard X-ray hardness increase, which is several days after the soft gamma-rays.  

Next, we investigated the spectral decomposition of the \(100-400\) keV flux during the observations to study the possible origin of the hard tail by looking at how the fluxes for the \texttt{CompTT} and \texttt{cutoff} components evolved.  The fluxes for the two components behaved quite differently with the \texttt{CompTT} flux decreasing until the source transitions to a soft state while the \texttt{cutoff} flux is at a constant level until approximately MJD 58016 when a short dip (\(< 0.5\) days) prior to the soft state.  

The Comptonization component is generally agreed to come from a corona while the origin of the hard tail is less clear.  The lack of correlation between the fluxes (Fig.~\ref{fig:flx_comps}) suggests that the origin of the high-energy component is different, which supports that it is related to the jet seen at optical/IR and radio wavelengths. 

A detection of polarized emission in soft gamma-rays would provide more constraints about the origin of the hard tail.  Unfortunately,  current instruments are unable to make such measurements for most sources.  Recently launched \textit{IXPE} will be able to detect polarization, but not above \(\sim 10\) keV \citep{2021arXiv211201269R}, and so it will not be able to study the high-energy component.  However, broadband observations may give some insight using capabilities already available.  In the case of a jet, the radio and IR emissions can provide interesting information to inform the high-energy observations since correlated evolutions are expected.  

Additionally, we investigated the spectral behavior of the rebrightening periods during the transition to the soft state.  Similar behavior was observed in MAXI J\(1820+070\) during a transition to the soft state.  Such features could possibly be explained by an expanding/contracting corona \citep{kara2019} or by material sent to the corona after disruptions of the inner accertion disk \citep{buisson2021}.  A better understanding of this behavior, seen only at hard X-rays, may provide insight of BH(C)'s during the state transition.

\begin{acknowledgments}
The research leading to these results has received funding from the European Union’s Horizon 2020 Programme under the AHEAD2020 project (grant agreement n. 871158).  J.R. acknowledges financial support under the INTEGRAL ASI/INAF No. 2019-35.HH.0. The \textit{INTEGRAL} SPI project has been completed under the responsibility and leadership of CNES.  We are grateful to ASI, CEA, CNES, DLR, ESA, INTA, NASA and OSTC for support.  This research has made use of MAXI data provided by RIKEN, JAXA and the MAXI team.  We thank the referee for the insightful comments and suggestions.  

\end{acknowledgments}

%





\end{document}